\documentclass[reprint,
superscriptaddress,
amsmath,amssymb,
 aps,
prb,
]{revtex4-1}

\usepackage{natbib}
\usepackage{graphicx}
\usepackage{dcolumn}
\usepackage{bm}
\usepackage{hyperref}

\def\permille{\ensuremath{{}^\text{o}\mkern-5mu/\mkern-3mu_\text{oo}}}
\setcitestyle{super,open={((},close={))}}

\begin{document}


\title{Self-formed, conducting LaAlO$_3$/SrTiO$_3$ micro-membranes}
\author{Alessia Sambri}
\affiliation{CNR-SPIN, Complesso Universitario di Monte S. Angelo, Via Cintia, 80126 Naples, Italy}
\email{catalexia@gmail.com}

\author{Mario Scuderi}
\affiliation{IMM-CNR, Strada VIII n. 5 Zona Industriale, 95121 Catania,Italy}

\author{Anita Guarino}
\affiliation{CNR-SPIN, Complesso Universitario di Monte S. Angelo, Via Cintia, 80126 Naples, Italy}
\affiliation{Department of Physical Sciences and Technologies of Matter, CNR-DSFTM, NFFA Trieste, Area Science Park - Basovizza Strada Statale 14, 34149 Trieste, Italy}

\author{Emiliano Di~Gennaro}
\affiliation{Dipartimento di Fisica ”E. Pancini”, Complesso Universitario di Monte S. Angelo, Via Cintia, 80126 Naples, Italy}
\affiliation{CNR-SPIN, Complesso Universitario di Monte S. Angelo, Via Cintia, 80126 Naples, Italy}

\author{Ricci~Erlandsen}
\affiliation{Niels Bohr Institutet, Universitetsparken 5, bygn. D, 2100 K$\o$benhavn, Denmark}

\author{Rasmus~T. Dahm}
\affiliation{Niels Bohr Institutet, Universitetsparken 5, bygn. D, 2100 K$\o$benhavn, Denmark}

\author{Anders~V. Bj$\o$rlig}
\affiliation{Niels Bohr Institutet, Universitetsparken 5, bygn. D, 2100 K$\o$benhavn, Denmark}

\author{Dennis~V.~Christensen}
\affiliation{Department of Energy Conversion and Storage, Technical University of Denmark, Roskilde, Denmark}

\author{Roberto Di~Capua}
\affiliation{Dipartimento di Fisica ”E. Pancini”, Complesso Universitario di Monte S. Angelo, Via Cintia, 80126 Naples, Italy}
\affiliation{CNR-SPIN, Complesso Universitario di Monte S. Angelo, Via Cintia, 80126 Naples, Italy}

\author{Umberto Scotti~di~Uccio}
\affiliation{Dipartimento di Fisica ”E. Pancini”, Complesso Universitario di Monte S. Angelo, Via Cintia, 80126 Naples, Italy}

\author{Salvatore Mirabella}
\affiliation{IMM-CNR, Strada VIII n. 5 Zona Industriale, 95121 Catania,Italy}
\affiliation{Dipartimento di Fisica e Astronomia,Università di Catania,Via S.Sofia 64,I-95123,Catania,Italy}

\author{Giuseppe Nicotra}
\affiliation{IMM-CNR, Strada VIII n. 5 Zona Industriale, 95121 Catania,Italy}

\author{Corrado Spinella}
\affiliation{IMM-CNR, Strada VIII n. 5 Zona Industriale, 95121 Catania,Italy}

\author{Thomas S.~Jespersen}
\affiliation{Niels Bohr Institutet, Universitetsparken 5, bygn. D, 2100 K$\o$benhavn, Denmark}

 \author{Fabio ~Miletto~Granozio}
\affiliation{Dipartimento di Fisica ”E. Pancini”, Complesso Universitario di Monte S. Angelo, Via Cintia, 80126 Naples, Italy}
\affiliation{CNR-SPIN, Complesso Universitario di Monte S. Angelo, Via Cintia, 80126 Naples, Italy}
\email{fabio.miletto@spin.cnr.it}

\begin{abstract}
Oxide heterostructures represent a unique playground for triggering the emergence of  novel electronic states and for implementing new device concepts. The discovery of 2D conductivity at the LaAlO$_3$/SrTiO$_3$ interface has been linking for over a decade two of the major current  research fields in Materials Science: correlated transition-metal-oxide systems and low-dimensional systems. A full merging of these two fields requires nevertheless the realization of LaAlO$_3$/SrTiO$_3$ heterostructures in the form of freestanding membranes. Here we show a completely new method for obtaining oxide hetero-membranes with micrometre lateral dimensions. Unlike traditional thin-film-based techniques developed for semiconductors and recently extended to oxides, the concept we demonstrate does not rely on any sacrificial layer and is based instead on pure strain engineering. We monitor through both real-time and post-deposition analyses, performed at different stages of growth, the strain relaxation mechanism leading to the spontaneous formation of curved hetero-membranes. Detailed transmission electron microscopy investigations show that the membranes are fully epitaxial and that their curvature results in a huge strain gradient, each of the layers showing a mixed compressive/tensile strain state. Electronic devices are fabricated by realizing ad hoc circuits for individual micro-membranes transferred on silicon chips. Our samples exhibit metallic conductivity and electrostatic field effect similar to 2D-electron systems in bulk heterostructures. Our results open a new path for adding oxide functionality into semiconductor electronics, potentially allowing for ultra-low voltage gating of a superconducting transistors, micromechanical control of the 2D electron gas mediated by ferroelectricity and flexoelectricity, and on-chip straintronics.
\end{abstract}

\maketitle

\section{Introduction}
The 2-dimensional electron system (2DES) forming at the LaAlO$_3$/SrTiO$_3$ (LAO/STO) interface is at the merging point of two major fields of research in condensed matter, i.e. correlated-electron and low-dimensional systems, and has been catalyzing the attention of a still-growing scientific community for 15 years \cite{ohtomo_high-mobility_2004}. Oxide-based 2DES have been demonstrated to display a number of outstanding functional properties, including high mobility \cite{ohtomo_high-mobility_2004,chen_high_mobility_2013},   superconductivity\cite{reyren_superconducting_2007, caviglia_electric_2008}, yet ill-understood magnetic properties\cite{brinkman_magnetic_2007,christensen_strain-tunable_2019} that can be stabilized by atomic engineering \cite{stornaiuolo_tunable_2016}, a strong Rashba type spin-orbit coupling \cite{caviglia_tunable_2010}, and an extraordinarily high charge-to-spin conversion efficiency \cite{lesne_highly_2016}. While the tunability of these properties by a gate potential is today largely established \cite{caviglia_electric_2008,caviglia_tunable_2010,stornaiuolo_tunable_2016,christensen_strain-tunable_2019,reiner_crystalline_2010}, it was not until very recently  that the possibility to tune their electronic properties by a mechanically induced flexoelectric potential was demonstrated \cite{zhang_modulating_electrical_2019}. This set of results makes oxide 2DES a candidate platform for a new generation of devices in the fields of electronics, spintronics, microelectromechanical systems and quantum technologies.

A new route bridging  the two realms of complex oxides and of low-dimensional system has been recently reported. By replicating concepts previously developed in semiconductor technology \cite{schmidt_thin_2001,huang_nanomechanical_2011,prinz_free-standing_2000,golod_fabrication_2001},  the realization of freestanding perovskite films was demonstrated resorting to water-soluble, epitaxial Sr$_3$Al${_2}$O${_6}$ sacrificial layer \cite{lu_synthesis_2016}.  This approach opens the perspective for application of methods and concepts developed in the field of 2D materials, as graphene, MoS$_2$ and related systems, to the arguably richest class of inorganic materials, i.e. transition metal oxides.  Among other epitaxial structures, crystalline freestanding ultrathin samples offer specific advantages, allowing the control over degrees of freedom as confinement, gating, curvature and strain that are only partially accessible in films locked on their substrate. Furthermore, analogously to the route successfully applied in electronics based on low dimensional Van der Waals materials\cite{geim_van_2013} or on bottom-up synthesized nanostructures\cite{nadj-perge_spinorbit_2010,larsen_semiconductor-nanowire-based_2015,mourik_signatures_2012,krogstrup_epitaxy_2015}, the availability of freestanding samples potentially allows to circumvent the problem of monolithical integration of oxides on semiconductors. The integration of devices based on epitaxial oxide  with semiconductor chips, in fact, has been addressed through the decades\cite{reiner_crystalline_2010,coll_towards_2019,McKee_crystalline_oxides_1998} 
with mixed success at a laboratory level and remains, in terms of industrial process compatibility, a partially unsolved challenge.  As a consequence, the method reported in \onlinecite{lu_synthesis_2016} has raised a major surge of interest and has been quickly replicated by other groups \cite{Hong_SciAdv_2017,Ji_2019_freestanding,Chen_freestanding_YBCO_2019,Birkholzer_Koster_New_Views_freestanding}. In spite of the many efforts on the topic,  a successful application to the realization of freestanding LAO/STO samples with high quality metallic properties has not been reported so far.

Here, we resort to strain engineering to demonstrate a totally independent approach, leading to the self-formation of freestanding epitaxial micro-heterostructures ($\mu$HSs) showing metallic conductivity down to cryogenic temperatures. By preserving the strained state of LAO and STO well above the critical thickness, we induce a devastating strain relaxation process, that fragments the surface in regularly-shaped LAO/STO $\mu$HSs. We analyze the breaking mechanism and characterize the individual $\mu$HSs in terms of curvature, microstructure, strain, strain gradient and transport. Finally, we discuss the perspectives opened by our findings both in term of fundamental understanding and of device applications.

\section{Strain relaxation in complementary growth regimes}

 Our LAO films were grown by pulsed laser deposition on TiO$_2$-terminated STO single crystals. We explored different deposition conditions, relying on previous experiences on plume spectroscopy and growth stoichiometry optimization \cite{sambri_plasma_2016,xu_impact_2013}, and fixing the O$_2$ background pressure to a relatively high value, $P_{O_2} = 2 \cdot 10^{-2}$ mbar. We ranged the film thickness up to about 180 nm, a value exceeding by almost two orders of magnitude the reported theoretical critical thickness h$_c$ for strain relaxation, confirmed by experiments on semiconductors growth (around 4 nm for 3\% mismatch, according to \onlinecite{people_calculation_1985}) and by about one order of magnitude a previously reported limit for pseudomorphic LAO/STO growth \cite{huijben_structureproperty_2009}. All samples went through an in-situ oxygen post-annealing at the end of the deposition. Further growth details are reported in the methods section. 

We now show that it is possible to control the film growth dynamics using the laser fluence on target as a tuning knob.  We select complementary growth regimes, imposing either a gradual or an abrupt strain relaxation mechanisms to the growing film. Similar complementary regimes, to some extent, were previously investigated by Choi et al  \cite{choi_atomic_2012}, that explored regions of the laser-fluence/background-oxygen-pressure phase diagram by varying the background pressure. 

\begin{figure}[ht!]
\centering
\includegraphics[width=0.95\linewidth]{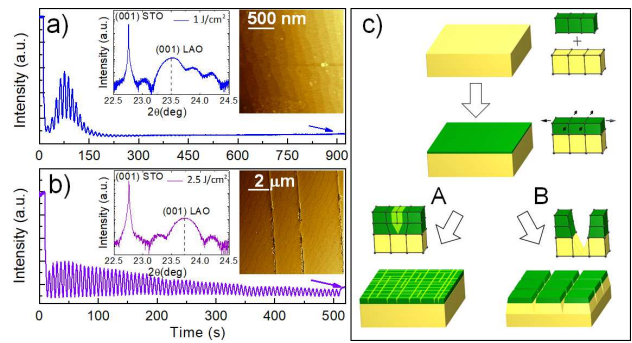}
\caption{(a) Specular (0,0) 
RHEED spot intensity evolution versus time during the growth of a 30 nm thick LAO sample grown in regime \textbf{A} (low fluence). The sample shows flat surfaces with regular terraces (AFM image in second inset) and a relaxed vertical lattice parameter  c = 0.378 nm as measured by x-ray diffraction (XRD, plot in first image);
(b) specular (0,0) RHEED spot intensity evolution versus time during the growth of a 30 nm thick LAO sample grown in regime \textbf{B} (high fluence). The samples show highly persistent RHEED oscillations, vertical surface cracks (AFM image in second inset), and a reduced vertical lattice parameter c = 0.374 nm consistent with tensile in-plain strain (XRD plot in first image).
(c) sketch showing the first three stages of the growth of LAO/STO samples in two different deposition regimes, at increasing thickness. The last stage in \textbf{B} regime, i.e. the formation of vertical cracks, starts occurring at about 25 nm.}
\label{fig:growth}
\end{figure}

In a low-laser-fluence growth regime, denoted as \textbf{A}, a gradual strain relaxation takes place in our samples through dislocation formation. Fig. \ref{fig:growth}a shows results from a 30 nm thick sample grown in this regime and confirms the growth of smooth conducting film, with terraced surfaces, damped reflection high-energy electron diffraction (RHEED) oscillations and a relaxed vertical lattice parameter. We find this kind of samples to be conducting  as shown in the Supplementary Information Fig. S1, up to a maximum investigated a thickness of 120 nm, well beyond the previously demonstrated thickness values \cite{choi_atomic_2012}. 

More interestingly, in a high-laser-fluence growth regime denoted as \textbf{B}, a complex evolution takes place as the film thickness increases.
The first three stages of this evolution are schematically shown in Fig. \ref{fig:growth}b, which partially anticipates results reported in the following. The films grow perfectly epitaxial, in a regime in which nucleation of dislocations is almost suppressed. Above a first critical thickness, vertical cracks appear at the surface, in agreement with mechanisms already observed in semiconductor film epitaxy \cite{lee_analysis_2006}. Such vertical cracks, leading to macroscopically insulating LAO/STO heterostructures, as checked  in Van der Pauw configuration, are observed in all samples with thickness of 30 nm or above, while all samples below 20 nm were conducting and crack free. Data collected on a 30 nm thick sample, reported in Fig. \ref{fig:growth}b, showed persistent RHEED oscillations, indicative of a steady growth regime persisting until the end of the deposition, a fully insulating behavior (data not reported), vertical surface cracks in atomic force microscopy (AFM, inset), scanning electron microscopy (SEM), optical images and TEM images (Supplementary Information, Figs. S3, S4 and S5) and a reduced vertical LAO lattice parameter c=0.374 nm (inset) consistent with tensile in-plane strain. The surface morphology, at this stage, presents similarities with cracked sample surfaces shown in \onlinecite{liu_epitaxial_2016} and \onlinecite{choi_atomic_2012}. The latter paper clarifies, in agreement with the conditions employed in this work, the role of high-oxygen-pressure growth in preventing strain relaxation through dislocations.

\begin{figure}[ht!]
\centering
\includegraphics[width=0.95\linewidth]{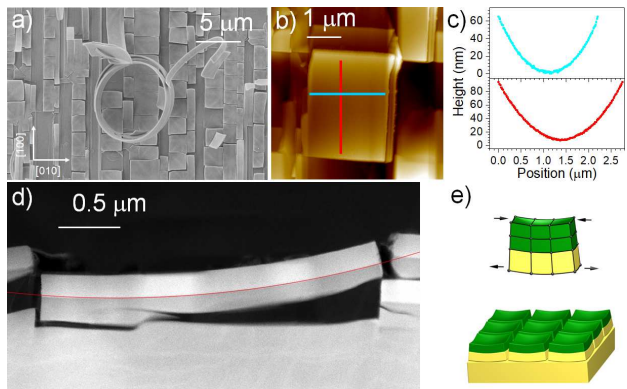}
\caption{(a) SEM picture of a typical surface of a sample in high fluence conditions at the fourth and final stage, showing membranes with typical lateral dimensions of 2-3 $\mu$m and, more rarely, tapes with lengths up to several tens of microns, rolled up in helices; (b) AFM image of a freestanding membrane of a 180 nm thick sample; c) AFM 1-D profiles collected on the red and cyan lines of the membrane shown in (b); (d) Low resolution cross-sectional TEM image of a typical membrane section in a 180 nm thick LAO film. The red arc following the interface belongs to a circle with radius R$_{curv}$ = 8 $\mu$m; (e) Sketch of the fourth and last stage of LAO/STO growth in our high fluence conditions: freestanding membranes break away from the substrate as a result of horizontal cracks formation.}
\label{fig:flakes}
\end{figure}

In Fig. \ref{fig:flakes}a, a SEM image of a 60 nm thick sample shows that the surface is covered by a mosaic of freestanding, regularly shaped, 2-3 $\mu$m sized membranes with sides aligned to the in-plane [100] and [010] directions of the substrate.
All membranes show a finite, similar, curvature, that is not expected for a pure LAO fragment. Instead, it is suggestive of a bilayer membrane made of materials in different strain state\cite{mourik_signatures_2012,krogstrup_epitaxy_2015,schmidt_thin_2001}. Some membranes keep from cracking along one direction up to a length exceeding 10 $\mu$m. When the length largely exceeds the curvature radius, membranes roll-up in helices, as in the case for the single with a curvature radius of about 5 $\mu$m is shown in Fig. \ref{fig:flakes}a. 
The AFM image of a 180 nm thick sample is shown in Fig. \ref{fig:flakes}b. The profiles extracted along orthogonal direction plotted in Fig. \ref{fig:flakes}c  are arch circles, though plotted with an expanded vertical scale, and allow to estimate a curvature radius of about 11 $\mu$m. 
A cross-view of a freestanding membrane section from the same sample is shown on the Z-contrast scanning transmission electron microscopy (STEM) micrograph of Fig. \ref{fig:flakes}d. This image shows that the curved membranes shown in Figs \ref{fig:flakes}b and c are actually bilayer heterostructures, reproducing at the microscale the intended macroscopic LAO/STO heterostructure. The measured LAO thickness (180 nm) agrees with the estimate from  RHEED oscillations (Supplementary Information, Fig. S2), and the STO layer thickness, though non-uniform, is grossly comparable. The estimated curvature radius is  8 $\mu$m. The image shown in the Supplementary Information, Fig. S5, demonstrates that most of the membranes are detached from the substrate, as the one shown in Fig. \ref{fig:flakes}d, and don't necessarily remain in proximity of the area where they were formed, while in some cases they are still partially connected to the underlying STO crystal . 

Data collected on different samples suggest that the thickness of the STO fragment stripped away from the substrate strongly correlates with LAO thickness. More statistics is needed to provide conclusive evidence and a quantitative analysis on this issue. Remarkably, no dislocations are found in membranes explored across their whole thickness: the growth of the LAO film appears to be fully coherent from the interface to the surface. 
All results from the previous panels are summarized by the sketch reported in Fig. \ref{fig:flakes}e, representing the fourth and final stage of the sequence shown in Fig. \ref{fig:growth}c for samples grown in regime \textbf{B}. As a consequence of the extreme strain energy accumulated during the perfect dislocation-free growth, curved free-standing bilayer membranes appear on the surface. Such $\mu$HSs self-form spontaneously, breaking away from the substrate in an abrupt strain relaxation process propagating a fracture plane parallel to the surface, within the STO single crystal substrate. No sacrificial layer is involved in this process and no chemical etching is needed. All data reported in the following part of this paper (Figs. \ref{fig:flakes}, \ref{fig:curvature} and \ref{fig:device}) refer to samples in such stage.

\section{Curvature, strain and strain gradient}
\begin{figure}[ht!]
\centering
\includegraphics[width=0.95\linewidth]{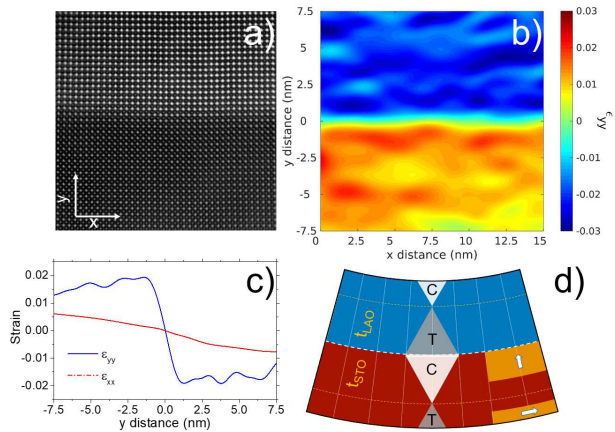}
\caption{(a) High-resolution Z-contrast STEM image showing the LAO/STO interface taken along the [100] direction showing that the macroscopic bending is reflected at the atomic scale in a perfect, dislocation-free, curved lattice; (b) corresponding $\epsilon_{yy}$ map, as obtained by GPA; (c) in-plane averaged $\epsilon_{xx}$ and ${\epsilon_{yy}}$ profiles vs. a growing (from STO to LAO) vertical coordinate; (d) geometric sketch of the heterostructure, showing the position of the neutral surfaces, at a distance t$_{STO}$ and t$_{LAO}$ from the interface, and the regions in tensile and compressive strain. The sections of the STO layers expected to show orthogonal ferroelectric order based on the STO strain phase diagram \cite{liu_epitaxial_2016} are identified by yellow areas}
\label{fig:curvature}
\end{figure}

 The high-resolution image and the data analysis reported in Fig. \ref{fig:curvature}a confirm that the macroscopically strained and bent state seen in the low resolution image in Fig. \ref{fig:flakes}d is coherently reproduced down to the atomic scale.  An accurate inspection of the image tells that the leftmost and rightmost planes perpendicular to the interface diverge from top to bottom, while the planes parallel to the interface present a faint curvature. This suggests that the system can be modelled as a dislocation-free, curved heterostructure,  coherently strained from the atomic to the micron scale.
 
 To quantitatively address the lattice deformations induced by such strained bent state state, we resorted to geometric phase analysis (GPA) \cite{hytch_quantitative_1998} starting from the high-resolution micrograph in Fig. \ref{fig:curvature}a. Here ${\epsilon_{xx}}$ and ${\epsilon_{yy}}$ are respectively defined as variations with respect to x-averaged in-plane (x direction, in the following) and out-of-plane (y direction)  lattice parameters at the interface of the heterostructure. Fig. \ref{fig:curvature}b shows the obtained $\epsilon_{yy}$  strain map from the corresponding high-resolution image in Fig. \ref{fig:curvature}a. This map shows that $\epsilon_{yy}$ is mostly dominated by the discontinuity of the out-of-plane parameter at the LAO/STO interface. 
 The ${\epsilon_{xx}}$ and ${\epsilon_{yy}}$ x-averaged profiles, plotted in Fig. \ref{fig:curvature}c, provide much finer information  about the lattice deformations induced in our heterostructure by the curved geometry. Beside the above mentioned $\epsilon_{yy}$ discontinuity at the interface, we observe in fact a negative slope of ${\epsilon_{xx}}$  in the positive y direction (i.e., from STO to LAO) and a positive slope of $\epsilon_{yy}$  away from the interface. As discussed in the Supplementary Information, Fig. S6, the slope of the ${\epsilon_{xx}}$ $(-0.95 \times 10^{-3} $nm$^{-1}$) vs. y plot equals, in absolute value, the inverse of the curvature radius $R_{curv}\sim 1\mu m$, while the slope of $\epsilon_{yy}$  on each side of the interface is related, to first order, to the Poisson ratio. 
 The sketch reported in Fig. \ref{fig:curvature}d summarizes our GPA results. It can be deduced by extrapolation that within each layer of our $\mu$HSs, a curved neutral surface is found, where the lattice (neglecting  the rhombohedral LAO distorsion) is cubic (${\epsilon_{yy}}={\epsilon_{yy}}$) and unstrained. The plots in Fig. \ref{fig:curvature}c suggest by extrapolation that such neutral surfaces ${\epsilon_{xx}}={\epsilon_{yy}}$ lie at about 10 nm from the interface, slightly outside the region thinned for high-resolution microscopy. The comparison between the GPA results and the predictions of a simple geometrical and mechanical model is discussed in the Supplementary Information, Sections S5 and S6.

The understanding of the equilibrium shape and strain distribution in freestanding epitaxial bilayer membranes has been addressed in some previous works based on semiconductors epitaxy \cite{mourik_signatures_2012,krogstrup_epitaxy_2015,schmidt_thin_2001,huang_nanomechanical_2011}. It can be mapped back to a classical continuous mechanics problem \cite{schmidt_thin_2001} first applied to bimetal thermostats \cite{timoshenko_analysis_1925}. To describe the strain state of our samples, we  address our $\mu$HSs as bilayer beams, rather than bilayer membranes. This approximation, valid when one dimension largely exceeds the other two, provides a direct comparison with our TEM images, due to the thinning of the samples along the zone axis, as well as with the helices. The membrane problem, not addressed here, is slightly more complex because the competition between curvatures along orthogonal directions slightly flattens the membranes. This justifies the bigger curvature radius measured in Fig. \ref{fig:flakes}b and c, with respect to Fig.  \ref{fig:flakes}d. 

Once the geometrical constraints set by the purely elastic deformation of our system are implemented in our model, the elastic energy minimization problem is reduced to finding the optimal curvature radius. Such problem is quantitatively addressed in Sections S6 and S7 of the Supplementary Information and an analytic solution is reported within the given approximations.  
At distances t$_{STO}$ and t$_{LAO}$ from the interface two zero-strain (``neutral'') arcs are found, where the strain vanishes and changes sign. Pure geometry tells that finding the optimal curvature radius is equivalent to finding the optimal distance of the neutral surfaces to the interface. We address here the specific case of a $\mu$HS composed of two layers with the same values of the respective Young moduli ($Y_{LAO}\sim Y_{STO}$) and thicknesses ($d_{LAO}\sim d_{STO}$).

The following equations apply:
\begin{align}
\label{eq1}
        t_{_{STO}}&=t_{_{LAO}}=\frac{d_{TOT}}{3}&\\ \
\label{eq2}
        R_{curv}&=\frac{2}{3}\frac{a_{ave}}{\Delta a}d_{TOT} 
  \end{align}
where R$_{curv}$ is the curvature radius of the interface, a$_{ave}$ is the average lattice parameter of the two layers and d$_{TOT}=d_{LAO}+d_{STO}$  is the total bilayer thickness. Eq. \ref{eq1}, Eq. \ref{eq2} can be considered as a specific case of Timoshenko formulas \cite{timoshenko_analysis_1925}, but they are also derived based on a simple geometrical argument in the the Supplementary Information, Sections S6 and S7. Numerical calculations based on formulas reported in Section S7 of the Supplementary Information show that the sample curvature is very stable in the surrounding of the chosen point ($Y_{LAO}= Y_{STO}$; $d_{LAO}=d_{STO}$): even considering a worst case scenario in which, e.g., both $d$ and $Y$ for one layer exceed by about $40\%$ the other, Eq. \ref{eq1}, Eq. \ref{eq2} remains accurate within less than $4\%$.

As for the Young moduli, we rely here on datasheets based on widespread ab-initio codes, mapping a broad subset of inorganic materials and providing, among other information, their elastic stiffness constants. As briefly discussed in the Supplementary Information, Section S4, the order of magnitude for $Y_{LAO}$ and $Y_{STO}$ is about 300 GPa, the LAO stiffness exceeding that of STO by no more than about 15$\%$. 

Considering the bulk values of LAO and STO lattice parameters
a value R$_{curv}\sim 7 \mu m $ is estimated for the sample in Fig.  \ref{fig:flakes}a, in reasonable agreement with the experimental value  R$_{curv}\sim 8 \mu m $ reported above. Eq. \ref{eq2} shows that decreasing R$_{curv}$ values are achieved at decreasing sample thickness, in agreement with our data. In particular, the much smaller R$_{curv}$ measured in Fig. \ref{fig:curvature} with respect to Fig. \ref{fig:flakes}d is related, though in a ill-defined geometry, to the further sample thinning procedure for high resolution electron microscopy.


The strain state of the self-formed micro-heterostructures is expected to have a large impact of the heterostructure properties. In particular, strain in STO can induce ferroelectricity even at room temperature \cite{haeni_room-temperature_2004} and enhance both the electron mobility \cite{Bharat_Stemmer_STO_Mobility_strain} and superconducting transition temperature \cite{Stemmer_2019_strain_supercond}. Yellow
areas in Fig. \ref{fig:curvature}d indicate regions where orthogonal ferroelectric polarizations (either parallel to the $x$ or to the $y$ coordinate) are expected to appear, possibly up to room temperature. The strain gradient in our heterostructures, namely the inverse of the curvature radius,  
is expected to induce the build-up of a giantic flexoelectric surface charge and potential, exceeding by orders of magnitude values obtained on bulk samples. Zhang et al. \cite{zhang_modulating_electrical_2019}, for example, resorting to dynamical bending of the macroscopic LAO/STO heterostructure, could apply a strain gradient not exceeding 1 m$^{-1}$. They showed the buildup of a flexoelectric potential of the order of 10mV and a consequent sizable variation of sheet resistance, carrier density and mobility of the 2DES. By comparison, the strain gradient measured in Fig. \ref{fig:curvature}c is about 1$\permille$ /nm (or  1 $\mu$m$^{-1}$), i.e., six orders of magnitude larger.



\section{Micro-heterostructures hosting a 2-dimensional electron system}
The crucial remaining question yet to be addressed, to assess the  interest of this system, is if our macroscopically insulating samples are indeed insulating down to the microscale, or are made of individually conducting but electrically disconnected $\mu$HSs. Two-terminal devices were fabricated by transferring $\mu$HSs to substrates of Si capped with 200 nm of SiO$_2$ gate insulator. 
\begin{figure}[ht!]
\centering
\includegraphics[width=0.95\linewidth]{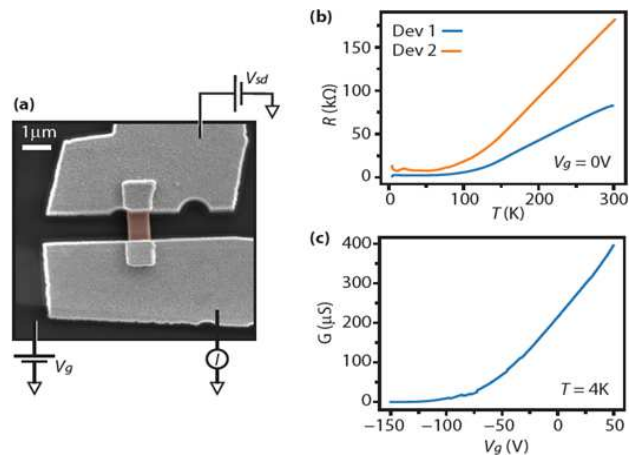}
\caption{(a) False color SEM micrograph of a typical $\mu$HS device after transfer to a Si/SiO$_2$ substrate and contacted in a two-terminal configuration. (b) Resistance as a function of temperature for two typical devices showing metallic behavior. (c) The gate dependence of the conductance of Device 1 at 4K. }
\label{fig:device}
\end{figure}

Individual $\mu$HSs were located and Ti/Au (300 nm/400nm) contacts were fabricated by electron-beam lithography (see Methods for further details). Figure \ref{fig:device}a shows a SEM micrograph of a typical device. Four device batches with a total of $\mathrm{\sim}$ 50 devices were fabricated and measured. A large spread in the room-temperature 2-contact resistance values were found,  most of them exceeding a value of 300 k$\Omega$ that generally increased to above the measurement limit upon cooling. We attribute this to dominating barriers at the metal $\mu$HSs interface as is commonly observed in mesoscopic semiconductor devices. Devices with $R\le 300k\Omega$ at room temperature, for which a good contact at the metal/2DES interface was achieved, showed metallic behavior upon cooling as shown in Fig. \ref{fig:device}b, and followed the trend expected for metallic LAO/STO 2DESs. Figure \ref{fig:device}c shows the two-terminal conductance upon varying the potential of the conducting Si substrate acting as a global back-gate through a composite barrier made of SiO$_2$ and of STO. The conductance increases with gate voltage V$_g$ as expected for a n-type 2DES again in agreement with expectation for macroscopic LAO/STO samples. Other metallic devices showed a weaker gate response, however with a consistent n-type trend.  Further work, including optimized contact recipes and the development of four-terminal devices, is needed to study the separate contributions of density and mobility.

\section{Summary and Perspectives}
In summary, we demonstrated the growth of self-formed, freestanding, epitaxial LAO/STO micro-heterostructures hosting a 2-dimensional electron system. Their curvature is analyzed from the micro to the atomic scale by resorting to several complementary microscopy techniques. Due to the extreme sensitivity of STO dielectric properties to strain, and to the giant strain gradient induced by this specific configuration, we argue that two orthogonal ferroelectric orientations might coexist in our samples at the distance of a few tens of nm and that huge flexoelectric effects are expected. Our micro-heterostructures are individually manipulated, transferred on Si/SiO$_2$ substrates, contacted and measured. Within the intrinsic limitations of the two-contact measurements performed so far, they show the standard transport properties of high-quality LAO/STO.

The potential fallout of this work is multifold. We demonstrate, in fact:

 a) A new strain-based technique allowing to grow highly crystalline, self-formed suspended membranes. This approach can be applied to other oxide, semiconductor or metal systems, circumventing the necessity to resort to a sacrificial layer. Prepatterned geometries based on trenches litographically defined on the STO surface prior to deposition can be employed to predetermine the geometry of the flakes.

b) The capability to create, manipulate and contact micro-sized LAO/STO heterostructures and potentially employ them as gate-tunable circuit components to transfer on substrates with pre-defined functionalities, such as microwave resonators and nanoscale gate arrays. This opens a route towards  integration of LAO/STO micro-heterostructure not only with semiconductors, but also with superconducting circuits, flexible electronics\cite{nomura_room-temperature_2004}, exfoliated 2D materials\cite{wang_growth_2017}, flexoelectric actuator\cite{bhaskar_flexoelectric_2016}, etc.; 

c) The ability to introduce in oxide science and technology novel degrees of freedom of major recent interest, i.e. curvature, strain and strain gradient. Application to the LAO/STO system opens the route to implementing devices of unprecedented complexity, in which curvature can tune the 2DES properties (in the normal or in the superconducting state) \cite{gentile_edge_2015,ying_tuning_2017}, either through direct geometric effects, or though variation of SrTiO$_3$ dielectric/ferroelectric/flexoeletric properties. We believe this to be the ideal system to study the interplay of strain, ferroelectricity \cite{haeni_room-temperature_2004}, superconducting transition temperature \cite{Stemmer_2019_strain_supercond}, flexoelectricity \cite{zhang_modulating_electrical_2019}, and domain boundary dynamics \cite{Kalisky_2017_domains}.

Our results enrich oxide 2DES science by providing access to many yet unexplored degrees of freedom and by bridging it to microscale materials engineering, micromechanics, ferro/flexoelectricity, mesoscopic/superconducting gate-tunable electronics and to the study of quantum and topological effects in curved 2D system.  

\section{Methods}
LAO films were grown by Pulsed Laser Deposition (PLD) on TiO$_2$-terminated (001) STO substrates held at 730$^\circ$C at oxygen background pressure of $2\cdot 10^{-2}$ mbar. A single crystal LAO target, mounted on a multi-target rotating carousel, was ablated by using a KrF excimer laser ($\lambda$ = 248 nm, pulse width= 20ns, ablation spot area = 0.78 mm$^2$, 45$^\circ$ angle of incidence) at a 3 Hz repetition rate. A mask allowed obtaining a homogeneous beam profile on the target surface. The target-substrate distance was fixed at 37 mm. After growth, an annealing at 500 $^\circ$C in 50 mbar of oxygen has been performed for 1h, before cooling down the samples in the same pressure condition. Regimes \textbf{A} and \textbf{B} only differ on the laser fluence, which are respectively in the ranges 1.0-1.5 and 2.0-2.5 $J/cm^{2}$.

LAO/STO samples topography and crystal structure have been characterized by means of AFM and XRD, respectively. The AFM images were acquired on an XE100 Park instrument operating in non-contact mode (amplitude modulation, silicon nitride cantilever from Nanosensor) at room temperature and in ambient conditions. The structural characteristics of the samples have been performed by means of a X'Pert MRD-PRO diffractometer in the high resolution configuration. A X-ray mirror and a Ge [220] monochromator were placed in the diffracted beam path to generate monochromatic Cu K$_{\alpha}$ X-rays, with $\lambda$= 1.54056 \AA. Crystal quality was determined by means of triple-axis attachment in which an additional three-bounce Ge (220) channel-cut analyzer was placed in front of the detector to obtain the same divergence of 12 arcsec as the incident beam. The diffractograms were acquired with a radiation of 40 kV and 30 mA, a step width of 0.001 $^\circ$C and an acquisition time of 10'' per step. 

Electrical characterization on macroscopic samples has been performed in a closed cycle cryostat down to 10K by means of a Keithley 2400 sourcemeter resorting a Van der Pauw configuration.

The LAO/STO composite membranes were prepared for a TEM side view observation by
embedding them in GATAN G1 epoxy glue and following with a conventional cross sectional TEM sample preparation. The sample was subsequently Ar ion milled using a Gatan Precision Ion Milling System with starting an energy of 3 keV, down to a final cleaning
energy of 200 eV. The structural analysis at the atomic level was performed with a (Cs)-probe-corrected JEOL JEM ARM200C operating in scanning TEM mode with a beam energy of 200 keV, using an high-angle annular detector resulting in ‘Z-contrast’ images. To take into account and correct both linear and non linear artifacts in experimental images due to unwanted motion of the electron probe with respect to the sample, we use a MATLAB implementation of the algorithm described in \onlinecite{ophus_correcting_2016}.
A commercial program, a Digital Micrograph
plug-in (DM 3.01 package, HREM Research
Inc.), was used for GPA. This algorithm reconstructs the displacement field by Fourier filtering two non-collinear Bragg vectors of the power spectrum generated from a high-resolution micrograph \cite{hytch_quantitative_1998}.

To fabricate electrical devices, the LAO/STO $\mu$HS were transferred from the growth substrate to a Si/SiO$_2$ substrate by gently touching the growth substrate with the corner of a clean-room paper and then the Si substrate (a similar technique is often applied for transfer of semiconductor nanostructures). Prior to transfer, the Si substrate was patterned with an array of metal alignment marks.  Suitable $\mu$HS were located with respect to the alignment grid using optical microscopy and electrical contacts in two-terminal configuration were defined by electron beam lithography using a $\sim$ 800 nm thick resist stack. To improve contact performance, a brief, gentle, Argon ion-milling was performed to induce conductivity in the exposed contact area before evaporation of Ti/Au (300 nm/400nm) contact metals. For both Argon milling and metal evaporation, a two-angle approach was employed with a shallow angle ($\sim$ 20 deg from substrate normal) from the directions normal to the current flow. This ensures Argon exposure and metal coverage of the edges of $\mu$HS which is where the actual contact is expected. 
It was confirmed on bulk STO samples that milling using the same parameters does not induce conductivity in the regions covered by resist.
Samples were bonded to a PCB sample holder and electrical measurements were performed using standard lock-in techniques employing a 100 $\mu$V ac excitation in a RT-4 K cryo-setup.

\bibliography{newbib}

\section*{Acknowledgements}
The authors acknowledge funding from the projects QUANTOX (QUANtum Technologies with 2D-OXides) of QuantERA ERA-NET Cofund in Quantum Technologies (Grant Agreement N. 731473) implemented within 10th European Union’s Horizon 2020 Programme and MIUR PRIN 2017 (Grant No. PRIN 20177SL7HC TOPSPIN). This work was also supported by nanoscience foundry and fine analysis (NFFA‐MIUR Italy Progetti Internazionali) project. The authors thank B. Della Ventura for help with optical measurements and A. Vecchione for the helpful discussion about XRD measurements.

\section*{Author contributions statement}
A.S. conceived the experiment and grew the samples with the help of A.G. and E.D.G., A.G. performed XRD analysis, E.D.G. performed the sheet resistance measurements on LAO/STO samples and optical measurements with the help of A.G., M.S. prepared the samples for TEM analysis, performed the measurements and developed the GPA analysis with the supervision of G.N. and C.S., R.E., R.T.D. and A.B. prepared the samples for the electrical measurements on LAO/STO micromembranes and performed the measurements with the supervision of T.J., D.V.C. contributed to experiments planning, data analysis and manuscript writing, R.D.C. performed the AFM measurements, S.M. performed the SEM measurements. F.M.G. supervised the study, coordinated the experiments by the different groups and performed the strain calculation with the help of R.D.C. and U.S.d.U..
A.S., E.D.G., and F.M.G. wrote the manuscript with inputs from all authors, and in particular from D.V.C and T.J. All authors discussed the results and contributed to their interpretation.
 
\section*{Competing interests}
The authors declare no competing interests

\section*{Additional information}
\noindent
\textbf{Supplementary information} is available for this paper \href{https://onlinelibrary.wiley.com/action/downloadSupplement?doi=10.1002\%2Fadfm.201909964&file=adfm201909964-sup-0001-SuppMat.pdf}{here}\\
\textbf{Correspondence and requests for materials} should be addressed to F.M.G. or A.S.\\

\end{document}